\documentclass[12pt, draftclsnofoot, onecolumn]{IEEEtran}

\usepackage{epstopdf}
\usepackage{url}
\usepackage{graphicx}
\usepackage{amsmath}
\usepackage{amsfonts}
\usepackage{amssymb, color}

\newtheorem{theorem}{Theorem}

\begin{document}

\title{Proportional Fairness in ALOHA Networks with RF Energy Harvesting }

\author{Zoran Hadzi-Velkov, Slavche Pejoski, Nikola Zlatanov, and Robert Schober

\thanks{Z. Hadzi-Velkov and S. Pejoski are with the Faculty of Electrical Engineering and Information Technologies, Ss. Cyril and Methodius University, 1000 Skopje, Macedonia (email: \{zoranhv, slavchep\}@feit.ukim.edu.mk).
N. Zlatanov is with the Department of Electrical and Computer Systems Engineering, Monash University, Clayton, VIC 3800, Australia (email: nikola.zlatanov@monash.edu).
R. Schober is with the Institute for Digital Communication, Friedrich-Alexander-University Erlangen-N\"urnberg, Erlangen D-91058, Germany (email: robert.schober@fau.de)}
\vspace{-9mm}
}

\markboth{ }{Shell \MakeLowercase{\textit{et al.}}: Bare Demo of IEEEtran.cls for Journals} \maketitle


\begin{abstract}
In this paper, we study wireless powered communication networks that employ the slotted ALOHA protocol, which is the preferred protocol for simple and uncoordinated networks. In the energy harvesting (EH) phase, the base station broadcasts radio frequency energy to the EH users (EHUs). The EHUs harvest the broadcasted energy and use it to transmit information back to the base station by contending for access to the uplink channel in the random access (RA) phase. In order to ensure fairness among the users, we propose a proportionally fair resource allocation scheme that exploits the RA nature of slotted ALOHA. Specifically, assuming statistical channel state information, we determine the optimal transmit power at the base station, the optimal durations of the EH and RA phases, the channel access probability, and the rate of each EHU.
\end{abstract}


\vspace{-10mm}

\section{Introduction}
The emergence of wireless powered communication networks (WPCNs) paves the way to perpetual and self-sustainable wireless networks [\ref{wpcn1}]-[\ref{lit1nova}]. The existing works on WPCNs (c.f. [\ref{wpcn2}] and references therein) typically assume channel sharing in the uplink based on time division multiple access (TDMA), which requires channel state information (CSI) and coordination among the energy harvesting (EH) users (EHUs) in order to align their successive transmissions and to optimize WPCN performance. On the other hand, a major application area for WPCNs is the Internet of Things (IoT), which is expected to comprise billions of low-cost and low-complexity wireless devices. Such nodes may be incapable of performing channel estimation and supporting the signalling required for mutual coordination. Therefore, random access schemes may be preferable over TDMA for uplink access as they do not require coordination nor full CSI.

Among the random access schemes, slotted ALOHA  [\ref{aloha1}] is often preferred because of its simplicity, resulting in its widespread use in both existing and emerging wireless networks, such as sensor networks, RFID systems, LTE systems [\ref{lit2nova}], and massive machine-type communication networks [\ref{lit3nova}]. Slotted ALOHA is also suitable for EH wireless networks, since EH users (EHUs) have a limited and sporadic energy supply and thus cannot utilize complex communication protocols. EH wireless networks based on slotted ALOHA have been studied in recent works. Specifically, the stability of the data queues of EH wireless networks was studied in [\ref{wpchaloha1}] and [\ref{wpcnaloha2}], whereas the tradeoff between the delivery probability and latency for these networks was studied in [\ref{wpcnaloha3}]. However, the application of the slotted ALOHA protocol in WPCNs has not been studied, yet.

WPCNs give rise to the {\it double near-far effect} [\ref{wpcn1}], which makes unfair resource utilization a much more severe problem in WPCNs compared to conventional wireless networks. ALOHA-based WPCNs can naturally tackle the fairness problem by allocating different (yet fixed) channel access probabilities to EHUs with different distances from the base station (BS). Specifically, EHUs having less energy would access the channel with a higher probability, whereas energy-rich EHUs would be granted a lower channel access probability. In this paper, we propose the application of the backlogged slotted ALOHA protocol [\ref{aloha1}] for random access in WPCNs, and optimize its parameters in order to guarantee {\it proportionally fair} (PF) resource allocation among the EHUs. PF resource allocation provides a certain level of fairness while avoiding bottlenecks created by weak users typical for other types of fairness, such as max-min fairness [\ref{pf}].

\vspace{-3mm}
\section{System Model}

\vspace{-2mm}
\subsection{Channel Model}
We consider a slotted ALOHA-based WPCN that consists of a BS and $K$ EHUs. Each network node is assumed to be equipped with a single antenna, and to operate in the half-duplex mode. We assume that each EHU has a rechargeable battery, which stores the RF energy harvested from the BS. The transmission time is divided into $M$ time slots of equal duration $T$. Each time slot is divided into two phases of fixed durations: an EH phase of duration $\tau_0 T$, and a random access (RA) phase of duration $(1-\tau_0)T$, where $\tau_0$ is a time-sharing parameter ($0 < \tau_0 < 1$). During the EH phase, the BS broadcasts RF energy to the EHUs at fixed output power $P_0$, and the EHUs harvest the transmitted energy. During the RA phase, the EHUs contend to access the channel to the BS for transmitting information at a fixed channel access probability. Let the channel access probability of the $k$th EHU be denoted by $q_k$, $1 \leq k \leq K$. Each EHU is assumed to always have information to send. In the RA phase of time slot $i$, the decision of the $k$th EHU whether to access the channel is determined by the outcome of a Bernoulli experiment, modeled by a random variable $I_k(i) \in \{0,1\}$, which is realized locally at the $k$th EHU. When $I_k(i) = 1$, the $k$th EHU transmits information at a fixed desired output power $P_{k0}$ and fixed rate $R_k$, and it is silent otherwise,
\begin{equation} \label{eq1}
I_{k}(i) = \begin{cases}
1, & \text{with probability } q_k \\
0, & \text{with probability } 1 - q_k ,
\end{cases}
\end{equation}
i.e., $E[I_k(i)] = q_k$ and $E[1 - I_k(i)] = 1 - q_k$, where $E[\cdot]$ denotes expectation. A collision occurs when two or more EHUs transmit information in the same RA phase.

The channel between the BS and the $k$th EHU ($1\leq k \leq K$) is modeled as a quasi-static block fading channel, where each fading block coincides with a single time slot. The channel fading is assumed to be a stationary and ergodic random process, whose instantaneous channel realizations follow the Nakagami-$m$ distribution. The Nakagami-$m$ channel model is general enough to accommodate the typical wireless fading environments of WPCNs. In time slot $i$, the fading power gains of the channel from the BS to the $k$th EHU and the channel from the $k$th EHU to the BS are denoted by $x_k(i)$ and $y_k(i)$, respectively. The average power gains and the fading parameters of both channels are assumed to be equal, i.e., $\Omega_{k}=E[x_k(i)]=E[y_k(i)]$ and $m_k=m_{Xk}=m_{Yk}$.

\vspace{-5mm}
\subsection{Energy Queue}
The rechargeable battery at the $k$th EHU is modeled as an energy queue with an infinite storage capacity.  Since the $k$th EHU harvests random amounts of energy that depend on the corresponding fading power gain, there is a chance that, in any given time slot, the $k$th EHU may not have sufficient energy stored in its battery to transmit with the desired output power $P_{k0}$. However, in [\ref{zlatanov}], it is proven under general conditions that, for $M \to \infty$, if an EHU with unlimited energy storage capacity employs a power allocation policy for which the average harvested energy is larger than or equal to the average amount of energy desired to be extracted from the battery, then this EHU can transmit with its desired output power in almost all time slots. More specifically, as $M \to \infty$, the number of time slots in which the battery cannot supply the $k$th EHU with the desired output power, $P_{k0}$, is negligible compared to the number of time slots in which the battery can provide the desired output power, $P_{k0}$. In this case, the considered EH network can be replaced with an equivalent non-EH network, where the average energy departure rate from the energy queue of the $k$th user, $E[P_{k0} I_k(i) (1-\tau_0)T]$, is less than or equal to the average energy arrival rate at the energy queue of the $k$th EHU, $E[\eta_k P_0 x_k(i) \tau_0 T]$, i.e., $E[P_{k0} (1 - \tau_0) I_k(i)] \leq E[\eta_k P_0 \tau_0 x_k(i)]$. Note that $\eta_k$ denotes the energy conversion efficiency of the $k$th EHU. The system performance is maximized by strict equality [\ref{zlatanov}], i.e.,
\begin{equation} \label{eq3}
\eta_k P_0 \tau_0 \Omega_{k} = P_{k0} (1 - \tau_0) q_k.
\end{equation}

\vspace{-5mm}
\subsection{Average Throughput}
For $M \to \infty$, the average throughput of the $k$th EHU, $\bar R_k$, is obtained as the product of the average rate of the user when the EHU successfully accesses the channel to the BS, $\hat R_k$, and the probability of successful channel access,
\begin{equation} \label{eq5}
\bar R_k = \hat R_k \, q_k \prod_{i \neq k} (1 - q_i).
\end{equation}
Assuming the $k$th EHU during the RA phase transmits at fixed transmission rate $(1-\tau_0)R_k$, the average rate $\hat R_k$ is determined by the product of the fixed rate and the probability of a non-outage event during the transmission of the $k$th EHU, i.e.,
\begin{eqnarray} \label{eq4}
\hat R_k &=& (1 - \tau_0) R_k \cdot \Pr \left \{\log_2 \left(1 + \frac{P_{k0} y_k(i)}{N_0} \right) \geq R_k \right\} \notag \\
&=& (1 - \tau_0) R_k \cdot \left(1 - F_{Y_k} \left(\frac{N_0 (2^{R_k} - 1)}{P_{k0}} \right) \right),
\end{eqnarray}
where $\Pr\{\cdot\}$ denotes probability, $F_{Y_k}(\cdot)$ is the cumulative distribution function (CDF) of the fading channel gain $y_k$, and $N_0$ is the additive white Gaussian noise (AWGN) power.

For Nakagami-$m$ fading, $\hat R_k$ in (\ref{eq4}) is given by
\begin{equation} \label{eq6}
\hat R_k = (1 - \tau_0) R_k \cdot \frac{1}{\Gamma(m_k)} \Gamma \left(m_k, \, \frac{m_k (2^{R_k} - 1) N_0}{P_{k0} \Omega_{k}} \right),
\end{equation}
where $\Gamma(m,x)$ is the incomplete Gamma function, defined as $\Gamma(m,x) = \int_{x}^{\infty} t^{m - 1} e^{-t}dt$, and $\Gamma(m)$ is the Gamma function.

\section{Proportionally Fair Resource Allocation }
When designing WPCNs, a crucial challenge is tackling the double near-far effect, which is much more severe then the near-far effect in traditional wireless networks. To achieve fairness in the system, we adopt the PF criterion for optimization [\ref{pf}], and maximize the sum of logarithms of the individual EHU throughputs. For the considered WPCN, PF resource allocation is achieved by joint optimization of $\tau_0$, $P_0$, $R_k$, $P_{k0}$, and $q_k$, $\forall k$, as the solution of the following maximization problem:
\begin{eqnarray}
\underset{\tau_0, P_0, P_{k0}, R_k, q_k} {\text{Maximize}} \ \sum_{k=1}^K \log \bar R_k \notag
\end{eqnarray}\vspace{-5mm}
\text{s.t.}
\vspace{-5mm}
\begin{eqnarray}\label{eq7}
\begin{array}{ll}
&C1: P_0 \leq P_{max} \\
&C2: P_0 \tau_0 \leq P_{avg}, \\
&C3: 0<\tau_0<1, \, 0<q_k<1, \,\, \forall k\\
&C4: \eta_k P_0 \tau_0 \Omega_{k} = P_{k0} (1-\tau_0)q_k, \,\, \forall k \\
\end{array}
\end{eqnarray}
where $\bar R_k$ is given by (\ref{eq5}), $C1$ is due to the maximum transmit power constraint at the BS, $P_{max}$, $C2$ is due to the average transmit power constraint at the BS, $P_{avg}$, and $C4$ is due to (\ref{eq3}). Using $C4$ in the objective function of (\ref{eq7}), the optimization variable $P_{k0}$ disappears from the objective function, and $\tau_0$, $P_0$, $R_k$, and $q_k$ are the remaining optimization variables. By combining (\ref{eq5})-(\ref{eq7}), we obtain
\begin{eqnarray} \label{eq8}
\underset{\tau_0, P_0, R_k, q_k} {\text{Maximize}} \ \sum_{k=1}^K \log \Big[(1-\tau_0) R_k \cdot q_k \prod_{i \neq k}(1-q_i) \Gamma \Big(m_k, \, \frac{(2^{R_k}-1) (1-\tau_0)q_k}{\tau_0 P_0 A_k} \Big) \Big] \notag \\
\text{subject to: } C1, C2, \text{ and } C3, \qquad \qquad \qquad \qquad \qquad \quad
\end{eqnarray}
where
\begin{equation} \label{eq9}
A_k = \frac{\eta_k \Omega_{k}^2}{m_k N_0}.
\end{equation}
Note that the double near-far effect is reflected by the term $\Omega_k^2$ in (\ref{eq9}).
The solution of (\ref{eq8}) is given in the following theorem.
\begin{theorem}\label{teorem1}
The optimal BS transmit power is given by
\begin{equation} \label{sol1}
P_0^* = P_{max}.
\end{equation}
The optimal duration of the EH phase is obtained as
\begin{equation} \label{sol2}
\tau_{0} =
\begin{cases}
\tau_0^*, & \text{if } 0 < \tau_0^* < \frac{P_{avg}}{P_{max}} \\
\frac{P_{avg}}{P_{max}}, & \text{if } \frac{P_{avg}}{P_{max}} \leq \tau_0^* < 1.
\end{cases}
\end{equation}
The optimal access probability of the $k$th EHU ($1 \leq k \leq K$) is given by
\begin{equation} \label{sol3}
q_{k} =
\begin{cases}
q_k^*, & \text{if } 0 < \tau_0^* < \frac{P_{avg}}{P_{max}} \\
q_k^{**}, & \text{if } \frac{P_{avg}}{P_{max}}  \leq \tau_0^* < 1.
\end{cases}
\end{equation}
The optimal rate of the $k$th EHU ($1 \leq k \leq K$) is given by
\begin{equation} \label{sol4}
R_k =
\begin{cases}
R_k^*, & \text{if } 0 < \tau_0^* < \frac{P_{avg}}{P_{max}} \\
R_k^{**}, & \text{if } \frac{P_{avg}}{P_{max}}  \leq \tau_0^* < 1.
\end{cases}
\end{equation}
The values of $\tau_0^*$ and $q_k^*, \forall k$ in (\ref{sol2}) and (\ref{sol3}) are determined as the solution of the following set of $K+1$ equations:
\begin{eqnarray} \label{a22}
\frac{1-K q_k^*}{1-q_k^*} = f_k \left(\frac{1-\tau_0^*}{\tau_0^*} \frac{q_k^*}{A_k  P_{max}} \left(-1 - \frac{1-q_k^*}{1-K q_k^*} \right. \right. \qquad \notag \\
\left. \left. \times \left[W \left(-\frac{1-q_k^*}{1-K q_k^*} \exp \left(- \frac{1-q_k^*}{1-K q_k^*} \right) \right) \right]^{-1} \right) \right), \,\, \forall k,
\end{eqnarray}
\begin{equation} \label{a13}
\tau_0^* = \frac{1}{K} \sum_{k=1}^K \frac{1-K q_k^*}{1-q_k^*},
\end{equation}
where $W(\cdot)$ is the Lambert-$W$ function, and  $f_k(\cdot)$ is an auxiliary function, defined as
\begin{equation} \label{a21}
f_k(x) = \frac{x^{m_k} \exp(-x)}{\Gamma(m_k,\, x)}.
\end{equation}
The value of $q_k^{**}, \forall k$ in (\ref{sol3}) is determined as the solution of (\ref{a22}) with $\tau_0$ replaced by $P_{avg}/P_{max}$. In (\ref{sol4}), the values of $R_k^*$ and $R_k^{**}$ are calculated as
\begin{eqnarray} \label{a18}
R_k = \frac{1}{\log(2)} \log \left(-\frac{1-q_k}{1-K q_k} \left[W \left(-\frac{1-q_k}{1-K q_k} \exp \left(-\frac{1-q_k}{1-K q_k} \right) \right) \right]^{-1} \right),
\end{eqnarray}
where $q_k$ is replaced by $q_k^*$ and $q_k^{**}$, respectively.

\end{theorem}

\begin{IEEEproof}
Please refer to Appendix A.
\end{IEEEproof}

\vspace{-3mm}

\section{Numerical Results}
In this section, we assume a WPCN with an even $K$, where one half (i.e., $K/2$) of the EHUs are placed on a circle of radius $r_1$ around the BS, and the other half (i.e., $K/2$) of the EHUs are on a concentric circle of radius $r_2$. Assuming the $k$th EHU is placed at distance $r_k$ from the BS, the deterministic path loss of the corresponding link to the BS is modeled as $\Omega_{k} = 10^{-3} r_k^{-3}$. Two EHU location sets are considered: $(r_1, r_2) = (10\text{m}, 20\text{m})$ (i.e., $\Omega_1/\Omega_2 = 8$) and $(r_1, r_2) = (10\text{m}, 12.5\text{m})$ (i.e., $\Omega_1/\Omega_2 = 2$). Without loss of generality, we assume $\eta_k  = 1, \forall k$. We also set $P_{max}=5$W, $P_{avg}=1$W, and $N_0 = 10^{-12}$W. The network sum throughput is calculated as $\sum_{k=1}^K \bar R_k$, whereas the system fairness is determined by the Jain's fairness index [\ref{jain}], as $(\sum_{k=1}^K \bar R_k)^2/(K \sum_{k=1}^K \bar R_k^2)$. Figs. 1 and 2 depict the two metrics resulting from the following resource allocation schemes: ($i$) the proposed scheme in Nakagami-$m$ fading with $m_k=3, \forall k$ (c.f. Theorem 1), ($ii$) the proposed scheme in the absence of fading (i.e., the static channel, c.f. Theorem 1 with $m_k \to \infty$), and ($iii$) a benchmark scheme in Nakagami-$m$ fading with $m_k=3, \forall k$.

In order to illustrate the benefits of optimizing the $q_k$s and $R_k$s according to Theorem 1, we have considered a benchmark slotted ALOHA scheme applied to the same WPCN topology, with identical values for all $q_k$s and $R_k$s, respectively. Specifically, we set $\tau_0=P_{avg}/P_{max}$, $q_k=1/K, \forall k$, and $R_k=R_0, \forall k$, where $R_0$ is the transmission rate that maximizes the average rate of an EHU located at distance $r_0=(r_1+r_2)/2$ from the BS $($i.e., $R_0$ maximizes (\ref{eq6}) when $\Omega_k=10^{-3} r_0^{-3}$$)$.

Figs. 1 and 2 show that both the network sum throughput and the fairness index grow with $K$. When the EHUs are distributed over a smaller range of distances from the BS (i.e., when $|r_2-r_1| = 2.5\text{m}$), the fairness level of the considered schemes is higher. Comparing the results for the Nakagami-$m$ channel with those for the static channel, we observe that channel fading deteriorates both performance metrics. However, the value of $m_k$ has less impact on the fairness than on the throughput. Compared to the benchmark scheme, the proposed scheme improves both performance metrics, where the improvement in throughput is more pronounced. Furthermore, the performance improvement of the proposed scheme over the benchmark scheme is more significant if the EHUs are distributed over a wider range of distances (i.e., when $|r_2-r_1| = 10\text{m}$). Overall, the proposed scheme ensures a high network sum throughput while guaranteeing fair resource sharing among the network nodes.

\begin{figure}[tbp]
\centering
\includegraphics[scale=0.55]{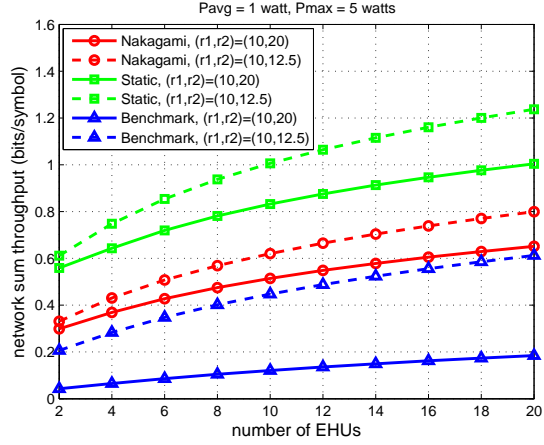} \vspace{-4mm}
\caption{Average throughput vs. number of EHUs } \vspace{-3mm}
\label{fig1}
\end{figure}

\begin{figure}[tbp]
\centering
\includegraphics[scale=0.55]{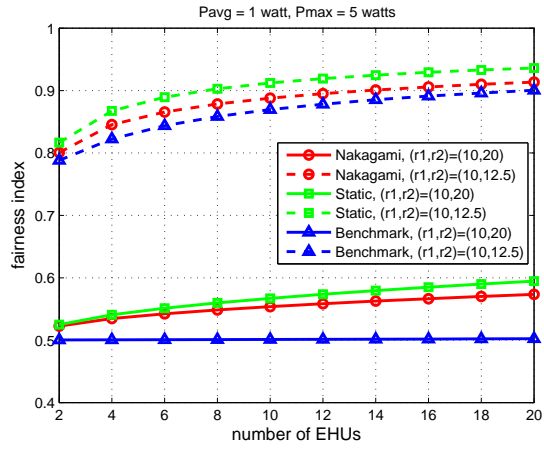} \vspace{-4mm}
\caption{Fairness index vs. number of EHUs } \vspace{-3mm}
\label{fig2}
\end{figure}

\vspace{-9mm}

\appendices
\section{Proof of Theorem 1}
First, we assume that constraint $C3$ is satisfied, whereas $C1$ and $C2$ are also satisfied but not with equality, i.e., $0<q_k<1$, $0<\tau_0<1$, $P_0<P_{max}$, and $P_0 \, \tau_0 < P_{avg}$. Under these assumptions, we check the properties of the objective function of (\ref{eq8}). Exploiting the properties of the $\log(\cdot)$ function, the objective function of (\ref{eq8}) is transformed as
\begin{eqnarray} \label{a1}
\mathcal L(P_0, R_k, q_k, \tau_0) = (K-1) \log (1 - q_k) + \sum_{k=1}^K \log \left(R_k (1-\tau_0) q_k \Gamma(m_k,X_k) \right),
\end{eqnarray}
where
\begin{equation} \label{a2}
X_k = \frac{2^{R_k} - 1}{C_k},
\end{equation}
and
\begin{equation} \label{a3}
C_k = \frac{\tau_0 P_0 A_k}{(1 - \tau_0)q_k}.
\end{equation}
Note that (\ref{a1}) is neither a concave nor a convex function. From the first derivative of (\ref{a1}) with respect to $P_0$, we obtain

\begin{equation} \label{a4}
\frac{d \mathcal L}{d P_0} = \frac{\tau_0}{1-\tau_0} \sum_{k = 1}^K  \frac{2^{R_k}-1}{C_k^2} \frac{A_k Z_k}{q_k} > 0,
\end{equation}
where
\begin{equation} \label{a5}
Z_k = -\frac{d}{dX_k} \log \left(\Gamma(m_k, X_k) \right) = \frac{X_k^{m_k-1} \exp(-X_k)}{\Gamma(m_k, X_k)}.
\end{equation}
From (\ref{a4}), we conclude that $\mathcal L(P_0, R_k, q_k, \tau_0)$ is an increasing function of $P_0$ for any arbitrary set of values for $\tau_0$, $R_k$, and $q_k$. Thus, in order to maximize $\mathcal L(P_0, R_k, q_k, \tau_0)$, $P_0$ should attain its maximum allowable value, which means that constraint $C1$ should be met with equality, i.e., $P_0^* = P_{max}$. Depending on whether $C2$ is satisfied with equality or not, we consider the following two cases: $P_{max} \tau_0 < P_{avg}$ (Case 1) and $P_{max} \tau_0 = P_{avg}$ (Case 2).

\vspace{-5mm}
\subsection{Case 1: $P_{max} \tau_0 < P_{avg}$}
Setting $P_0 = P_{max}$, the function $\mathcal L(P_{max}, R_k, q_k, \tau_0)$ should be maximized with respect to $R_k$, $q_k$, and $\tau_0$. Clearly, the solution is at one of the critical points of $\mathcal L$, which are found by setting the partial derivatives of the objective function with respect to $R_k$, $q_k$, and $\tau_0$ to zero. A critical point can be a local maximum, a local minimum, or an inflection point. In what follows, we will show that the objective function of (\ref{eq8}) has a single critical point, which is a global maximum. By setting the first derivatives of $\mathcal L(P_{max}, R_k, q_k, \tau_0)$ with respect to $R_k$, $q_k$, and $\tau_0$ to zero, we obtain:
\begin{align}
& \frac{d \mathcal L}{dR_k} = \frac{1}{R_k} - \frac{Z_k 2^{R_k} \log(2)}{C_k} = 0, \,\, \forall k, \label{a7} \\
& \frac{d \mathcal L}{dq_k} = \frac{1}{q_k} - \frac{K-1}{1-q_k} - \frac{\tau_0 A_k Z_k (2^{R_k}-1)}{(1-\tau_0) C_k^2 \, q_k^2} = 0, \,\, \forall k, \label{a8} \\
& \frac{d \mathcal L}{d\tau_0} = -\frac{K}{1-\tau_0} + \sum_{k=1}^K \frac{A_k Z_k (2^{R_k}-1)}{(1-\tau_0)^2 C_k^2 \, q_k} = 0. \label{a9}
\end{align}
Inserting (\ref{a5}) into (\ref{a7}) leads to
\begin{eqnarray} \label{a19}
\left(\frac{1-2^{R_k}}{2^{R_k} C_k} \right)^{m_k-1} \exp \left(\frac{2^{R_k}-1}{2^{R_k} C_k} \right)
= -\frac{2^{R_k} C_k \log(2)}{R_k} \,\, \Gamma \left(m_k, \frac{1-2^{R_k}}{2^{R_k} C_k} \right),
\end{eqnarray}
which, using (\ref{a21}) and some reordering, is further transformed into (\ref{a22}). Combining (\ref{a8}) and (\ref{a9}), we arrive at (\ref{a13}). Furthermore, (\ref{a8}) can be solved in closed form as (\ref{a18}).

\vspace{-6mm}
\subsubsection{Uniqueness of solution of the set of equations (\ref{a22}), (\ref{a13})} According to (\ref{a18}), a positive value of $R_k$ necessitates $B_k > 1$, or equivalently, $0<q_k<1/K$. For any given $\tau_0$ in the interval $(0, 1)$, there is a single solution $q_k^0$ of (\ref{a22}) in the interval $(0, 1/K)$, because the function on the left hand side of (\ref{a22}) is strictly decreasing from 1 to zero, whereas the right hand side is strictly increasing from 0 to $\infty$. Let $q_k^0 = g(\tau_0)$ denote the dependence of $q_k^0$ on $\tau_0$. Clearly, $q_k^0$ is increasing for $\tau_0 \in (0, 1)$, and its inverse $\tau_0 = g^{-1}(q_k)$ also is increasing for $q_k \in (0, 1/K)$ according to (\ref{a22}). On the other hand, according to (\ref{a13}), $\tau_0 = (1/K) \sum_{k=1}^K (1-K q_k)/(1-q_k) \triangleq h(q_k)$ is decreasing for $q_k \in (0, 1/K)$. Thus, there is a single solution $q_k^*$ of the equation $h(q_k) = g^{-1}(q_k)$ in the interval $(0, 1/K)$. This single solution corresponds to a global maximum since the objective function of (\ref{eq8}) at $q_k=0$ and $q_k = 1$ is $-\infty$.

\vspace{-7mm}
\subsection{Case 2: $P_{max} \tau_0 = P_{avg}$}
In this case, the optimal value of $\tau_0$ is $\tau_0 = P_{avg}/P_{max}$. Therefore, we now aim to maximize $\mathcal L \left(P_{max}, R_k, q_k, \frac{P_{avg}}{P_{max}} \right)$ with respect to $q_k$ and $R_k$. Following similar steps as for Case 1, the optimal value of $q_k, \forall k$, is obtained as in (\ref{a22}) with $\tau_0$ replaced by $P_{avg}/P_{max}$.

\end{document}